\documentstyle[11pt,paspconf,epsf]{article}
\begin{document}
\title{Intermittency as a possible underlying mechanism for solar and
stellar variability}
\author{Reza Tavakol \& Eurico Covas}
\affil{
Astronomy Unit\\
School of Mathematical Sciences \\
Queen Mary \& Westfield College \\
Mile End Road \\
London E1 4NS, UK
}
\begin{abstract}
We briefly discuss the status of the {\it intermittency hypothesis},
according to which the grand minima type variability
in solar-type stars may be
understood in terms of dynamical intermittency.
We review concrete examples which establish this
hypothesis in the mean-field setting.
We discuss some difficulties and open problems
regarding the establishment of this hypothesis in
more realistic settings as well as its
operationally decidability.
\end{abstract}
\section{Introduction}
It is now well established that middle-aged solar-type stars show variability
on a wide range of time scales, including the intermediate time scales of
$\sim 10^0 - 10^4$ years (Weiss 1990). The evidence for the latter comes from
a variety of sources, including observational, historical and proxy records.
Many solar-type stars seem to show cyclic types of behaviour in their
mean magnetic fields (e.g. Weiss 1994, Wilson 1994), which in the case of the
Sun have a period of nearly 22 years. Furthermore, the studies of the
historical records of the annual mean sunspot data since 1607 AD show the
occurrence of epochs of suppressed sunspot activity, such as the {\it Maunder
minimum} (Eddy 1976, Foukal 1990, Wilson 1994, Ribes \& Nesme-Ribes 1993,
Hoyt \& Schatten 1996). Further research, employing $^{14}C$ (Eddy 1980,
Stuiver \& Quey 1980, Stuiver \& Braziunas 1988, 1989) and $^{10} B$ (Beer
{\em et al.} 1990, 1994a,b, Weiss \& Tobias 1997) as proxy indicators, has
provided strong evidence that the occurrence of such epochs of reduced
activity (referred to as {\it grand minima}) has persisted in the past with
similar time scales, albeit irregularly.

These latter, seemingly irregular, variations are important for two reasons.
Firstly, the absence of naturally occurring mechanisms in solar and stellar
settings, with appropriate time scales (Gough, 1990), makes the explanation of
such variations theoretically challenging.  Secondly, the time scales of such
variations makes them of potential significance in understanding the climatic
variability on similar time scales (e.g.\ Friis-Christensen \& Lassen 1991,
Beer {\em at al.} 1994b, Lean 1994, Stuiver, Grootes \& Braziunas 1995,
O'Brien {\em et al.} 1995, Baliunas \& Soon 1995, Butler \& Johnston 1996,
White {\em et al.} 1997).
In view of this, a great deal of effort has gone into trying to understand the
mechanism(s) underlying such variations by employing a variety of approaches.

Our aim here is to give a brief account of some recent results that may
throw some new light on our understanding of such variations.

\section{Theoretical frameworks}
Theoretically there are essentially two frameworks within which such
variabilities could be studied: stochastic and deterministic.

Here we mainly concentrate on the deterministic approach and recall that given
the usual length and nature of the solar and stellar observational data, it is
in practice difficult to distinguish between these two frameworks (Weiss
1990). Nevertheless, even if the stochastic features play a significant role
in producing such variations, the deterministic components will still be
present and are likely to play an important role.

The original attempts at understanding such variabilities were made within the
linear theoretical framework. An important example is that of linear
mean-field dynamo models (Krause \& R\"adler 1980)
which succeeded in reproducing the nearly 22 year cyclic behaviour.
Unfortunately such linear models cannot easily and naturally\footnote{It is
worth bearing in mind that one can always produce complicated looking
behaviour within the linear framework, by combining many simpler behaviours.
The crucial point is that in this case complexity in behaviour requires a
complicated underlying mechanism. Furthermore, there are qualitative
differences, in terms of spectra and other dynamical indicators, between
complicated dynamical behaviours produced by linearly complex and nonlinearly
chaotic systems.} account for the complicated, irregular looking solar
and stellar variability.

The developments in nonlinear dynamical systems theory, over the last few
decades, have provided an alternative framework for understanding such
variability. Within this nonlinear deterministic framework, irregularities of
the grand minima type are probably best understood in terms of various types
of dynamical intermittency, characterised by different statistics over
different intervals of time. The idea that some type of dynamical
intermittency may be responsible for understanding the Maunder minima type
variability in the sunspot record goes back at least to the late 1970's (e.g.
Tavakol 1978, Ruzmaikin 1981, Zeldovich {\em et al.} 1983, Weiss {\em et al.}
1984, Spiegel 1985, Feudel {\em et al.} 1994). We shall refer to the
assumption that grand minima type variability in solar-type starts can be
understood in terms of some type of dynamical intermittency as the {\it
intermittency hypothesis}.

To test this hypothesis one can proceed by adopting either a quantitative or a
quantitative approach.

\subsection{Quantitative approach}
Given the complexity of the underlying equations, the most direct approach to
the study of dynamo equations is numerical. Ideally one would like to
start with the full 3--D dynamo models with the least number of simplifying
assumptions and approximations. There have been a great deal of effort in this
direction over the last two decades (e.g. Gilman 1983, Nordlund {\em et al.}
1992, Brandenburg {\em et al.} 1996, Tobias 1998). The difficulty of dealing
with small scale turbulence has meant that a detailed fully self-consistent
model is beyond the range of the computational resources currently available,
although important attempts have been made to understand turbulent dynamos in
stars (e.g. Cattaneo, Hughes \& Weiss 1991, Nordlund {\em et al.} 1992, Moss
{\em et al.} 1995, Brandenburg {\em et al.} 1996, Cattaneo \& Hughes 1996) and
accretion discs (e.g. Brandenburg {\em et al.} 1995, Hawley {\em et al.}
1996). Such studies have had to be restricted to the geometry of a Cartesian
box, which in essence makes them local dynamos, whereas magnetic fields in
astrophysical objects are observed to exhibit large scale structure, related
to the shape of the object, and thus can only be captured fully by global
dynamo models (Tobias 1998). Furthermore, despite great advancements in
numerical capabilities, these models still involve approximations and
parametrisations and are extremely expensive numerically, specially if the aim
is to make a comprehensive search for possible ranges of dynamical modes of
behaviours as a function of control parameters\footnote{Which at times would
require extremely long runs to transcend transients.}.

An alternative approach, which is much cheaper numerically, has been to employ 
mean-field dynamo models. Despite their idealised nature, these models reproduce
some features of more complicated models and allow us to analyse certain
global properties of magnetic fields in the Sun. For example, the dependence
of various outcomes of these models (such as parity, time dependence, cycle
period, etc.) on global properties, including boundary conditions, have been
shown to be remarkably similar to those produced by full three-dimensional
simulations of turbulent models (Brandenburg 1999a,b). This gives some
motivation for using these models for our studies below.

A number of attempts have recently been made to numerically
study such models, or their
truncations, to see whether they are capable of producing the
grand minima type behaviours.  There are a number of problems with these
attempts. Firstly, the developments in dynamical systems theory over the
last two decades have uncovered a number of theoretical mechanisms for
intermittency, each with their dynamical and statistical signatures.
Secondly, the simplifications and approximations involved in these models,
make it difficult to decide whether a particular type of behaviour obtained
in a specific model is in fact generic. And finally, the characterisation of
such numerically obtained behaviours as ``intermittent'' is often
phenomenological and based on simple observations of the resulting time
series (e.g. Zeldovich {\em et al.} 1983, Jones {\em et al.} 1985, Schmalz
\& Stix 1991, Feudel {\em et al.} 1993, Covas {\em et al.} 1997a,b,c,
Tworkowski {\em et al.} 1998, and references therein), rather than a
concrete dynamical understanding coupled with measurements of the predicted
dynamical signatures and scalings. There are, however, examples where
the presence of various forms of intermittency has
been established concretely
in such dynamo models, by
using various signatures and scalings (Brooke 1997,
(Covas \& Tavakol 1997, Covas {\em et al.} 1997c, Brooke {\em et al.} 1998,
Covas \& Tavakol 1998, Covas {\em et al.} 1999b).

\subsection{Qualitative approach}

Given the inevitable approximations and simplifications involved
in dynamo modelling (specially given the turbulent nature of the
regimes underlying such dynamo behaviours and hence the parametrisations
necessary for their modelling in practice), a great deal of effort has
recently gone into the development of approaches that are in some sense
generic. The main idea is to start with various qualitative features that
are thought to be commonly present in such settings and then to study
the generic dynamical consequences of such
assumptions.

Such attempts essentially fall into the following categories. Firstly, there
are the low dimensional ODE models that are obtained using the Normal Form
approach (Spiegel 1994, Tobias {\em et al.} 1995, Knobloch {\em et al.} 1996).
These models are robust and have been successful in
accounting for certain aspects of the dynamos,
such as several types
of amplitude modulation of the magnetic field
energy, with potential relevance for solar variability
of the Maunder minima type.

The other approach is to single out the main generic ingredients of such
models and to study their dynamical consequences. For axisymmetric dynamo
models, these ingredients consist of the presence of invariant subspaces,
non-normal parameters and non-skew property. The dynamics underlying such
systems has recently been studied in (Covas {\em et al.}, 1997c,1999b; Ashwin
{\em et al.} 1999). This has led to a number of novel
phenomena, including a new type of intermittency, referred to as
{\it in--out intermittency},
which we shall briefly discuss in section
\ref{intermittencies}

\section{Models}

The standard mean-field dynamo equation is given by
\begin{equation} \label{dynamo}
\frac{\partial {\bf B}}{\partial t}=
\nabla \times \left( {\bf u} \times {\bf B} + \alpha {\bf B} - \eta_t
\nabla \times {\bf B} \right),
\end{equation}
where ${\bf B}$ and ${\bf u}$ are the mean magnetic field and mean velocity
respectively and the turbulent magnetic diffusivity $\eta_t$ and the
coefficient $\alpha$ arise from the correlation of small scale turbulent
velocities and magnetic fields (Krause \& R\"adler, 1980). In axisymmetric
geometry, eq.\ (1) is solved by splitting the magnetic field into meridional
and azimuthal components, ${\bf B} = {\bf B_{p}} + {\bf B_{\phi}}$, and
expressing these components in terms of scalar field functions {${\bf B_{p}} =
\nabla \times A \hat{\phi}$, ${\bf B}_{\phi} = B \hat{\phi}$}.

In the following we shall also employ a family of truncations of the one
dimensional version of equation (\ref{dynamo}), along with a time dependent
form of $\alpha$, obtained by using a spectral expansions of the form:
\begin{eqnarray}
\frac{dA_n}{dt}&=&-n^2A_n+\frac{D}{2}(B_{n-1}+B_{n+1})+\nonumber
\sum_{m=1}^{N}\sum_{l=1}^{N}{\cal F}(n,m,l)B_mC_l,\nonumber\\
\frac{dB_n}{dt}&=&-n^2B_n+\sum_{m=1}^{N}{\cal G}(n,m)A_m,\label{truncated} \\
\nonumber\frac{dC_n}{dt}&=&-\nu n^2 C_n
-\sum_{m=1}^{N}\sum_{l=1}^{N}{\cal H}(n,m,l)A_mB_l.
\end{eqnarray}
where $A_n$, $B_n$ and $C_n$ are derived from the spectral expansion of the
magnetic field ${\bf B}$ and $\alpha$ respectively, ${\cal F, H}$ and ${\cal
G}$ are coefficients expressible in terms of $m,n$ and $l$, $N$ is the
truncation order, $D$ is the dynamo number and $\nu$ is the Prandtl number
(see Covas {\em et al.} 1997a,b,c for details).

\section{Different forms of intermittency in ODE and PDE dynamo models}
\label{intermittencies}
Recent detailed studies of axisymmetric mean field 
dynamo models have produced concrete evidence for the
presence of various forms of dynamical intermittency in such models. We shall give a
brief overview of these results in this section.

\begin{figure}[!htb]
\centerline{\def\epsfsize#1#2{0.6#1}
\epsffile{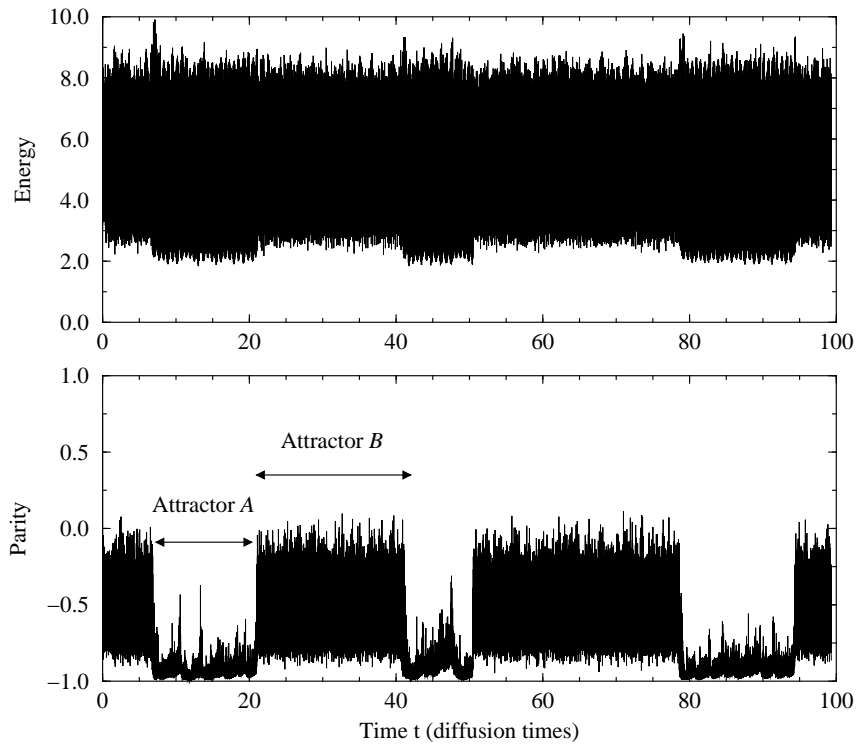}}
\caption{\label{crisis1}
Example of crisis induced intermittency in a shell dynamo with a cut,
with $r_0=0.2$,
$C_\alpha=25.5$, $C_\Omega=-10^4$, $\theta_0=45^{\circ}$.
See Covas {\em et al.}, 1999a for details of the model.}
\end{figure}

\subsection{Crisis (or attractor merging) intermittency}

A particular form of this type of intermittency, discovered by Grebogi, Ott \& Yorke
(Grebogi {\em et al.} 1982, 1987), is the so called ``attractor merging
crisis'', where as a system parameter is varied, two or more chaotic
attractors merge to form a single attractor. There is both experimental and
numerical evidence for this type of intermittency (see for example Ott (1993)
and references therein). We have found
concrete evidence for the presence of such a behaviour
in a 6-dimensional truncation of mean-field dynamo model of the
type (\ref{truncated}) (Covas \& Tavakol 1997) and more recently, 
in a PDE model of type (\ref{dynamo}) (see Covas \& Tavakol
(1999) for details). Fig.\ \ref{crisis1} shows an example of the latter
which clearly demonstrates the merging of two attractors, with
different time averages for energy and parity.
For a concrete characterisation and scaling,
see Covas \&
Tavakol (1999).

\subsection{Type I-Intermittency}
\begin{figure}[!htb]
\centerline{\def\epsfsize#1#2{0.6#1}
\epsffile{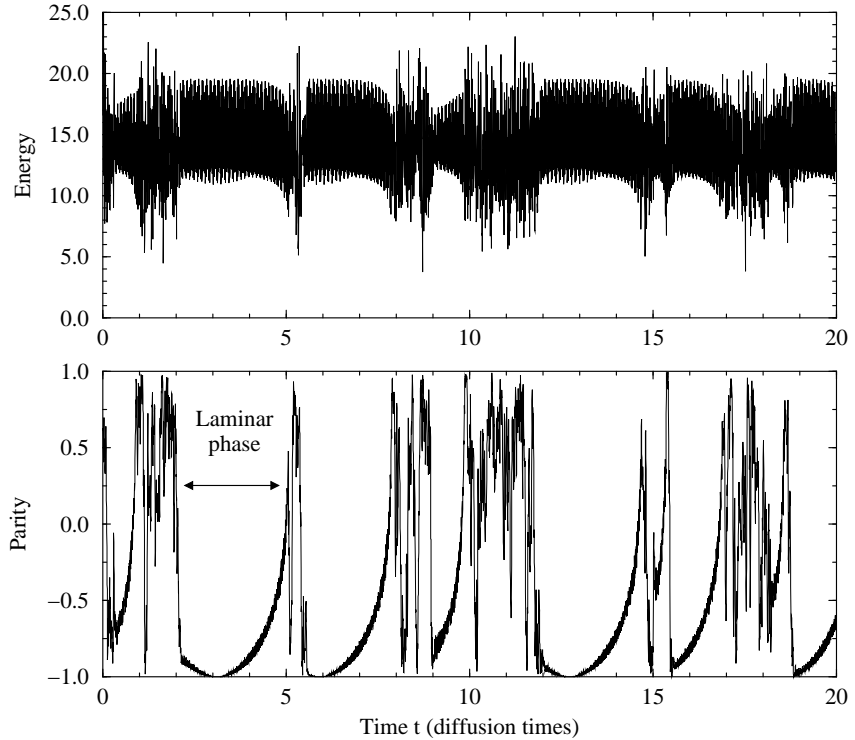}}
\caption{\label{typeIa}
Example of Type-I intermittency in a shell dynamo with a cut, with $r_0=0.7$,
$C_\alpha=28.0$, $C_\Omega=-10^4$, $\theta_0=45^{\circ}$.
See Covas {\em et al.}, 1999a for details of the model.
}
\end{figure}

This form of intermittency, first discovered by Pomeau and Manneville in the
early 1980's (Pomeau \& Manneville 1980), has been extensively studied
analytically, numerically and experimentally (see Bussac \& Meunier 1982,
Richter {\em et al.} 1994 and references therein). It is identified
by long almost regular phases interspersed by
(usually) shorter chaotic bursts.
In particular, this type of intermittency has been
found in a 12--D truncated dynamo model of type (\ref{truncated})
(Covas {\em et al.} 1997c), and more
recently in a PDE dynamo model of type (\ref{dynamo}) (Covas \& Tavakol
1999). Fig.\ \ref{typeIa} gives an example of such time series, where the
irregular interruptions of the laminar phases by chaotic bursts can easily be
seen. For a concrete characterisation, including the scaling for the average
length of laminar phases see Covas \& Tavakol (1999).

\subsection{On-Off and In-Out Intermittency}
An important feature of systems with symmetry (as in the case of solar and
stellar dynamos) is the presence of invariant submanifolds. It may happen that
attractors in such invariant submanifolds may become unstable in transverse
directions. When this happens, one possible outcome could be that the
trajectories can come arbitrarily close to this submanifold but also have
intermittent large deviations from it. This form of intermittency is referred
as {\it on-off intermittency} (Platt {\em et al.} 1993a,b). Examples of this type
of intermittency have been found in dynamo models, both phenomenologically
(Schmitt {\em et al.}, 1996) and concretely in truncated dynamo models of the
type (\ref{truncated}) (Covas {\em et al.} 1997c).

A generalisation of on-off intermittency, the in-out
intermittency, discovered recently (Ashwin {\em et al.} 1999)
is expected to be generic for axisymmetric dynamo settings.
The crucial distinguishing feature of this type
of intermittency is that, as
opposed to on-off intermittency, there can be different invariant sets
associated with the transverse attraction and repulsion to the invariant
submanifold, which are not necessarily chaotic. This gives rise to
identifiable signatures and scalings (Ashwin {\em et al.} 1999).

Concrete evidence for the occurrence of this type of intermittency has been
found recently in both PDE and truncated dynamo models of the types
(\ref{dynamo}) and (\ref{truncated}) respectively (see Covas {\em et al.}
(1999a,b) for details).

\section{Intermittency hypothesis: theory and observation}
In the previous section, we 
have summarised concrete evidence for the presence of four different types of
dynamical intermittency in both truncated and PDE mean-field dynamo models.
>From a theoretical point of view,
the intermittency hypothesis may therefore be said to have been
established, at least within this family of mean-field models. What remains
to be seen is whether these types of intermittency
still persist in more realistic models. 
An encouraging development in this connection
is the discovery of a type of intermittency
which is expected to occur generically
in axisymmetric dynamo settings, independently
of the details of specific models. 
Despite these developments,
testing the
intermittency hypothesis poses a number of difficulties in practice:

\begin{enumerate}
\item
Observationally, all precise dynamical characterisation of solar and stellar
variability are constrained by the length and the quality of the available
observational data. This is particularly true of the intermediate (and of
course longer) time scale variations. Such a characterisation is further
constrained by the fact that some of the indicators of such
mechanisms, such as scalings, require very long
and high quality data.

\item
Theoretically, there is now a large number of such mechanisms, some of which
share similar signatures and scalings, which could potentially
complicate the process of differentiation
between the different mechanisms.

\item
An important feature of real dynamo settings is the inevitable presence
of noise. This calls for a theoretical and numerical study of effects of noise
on the dynamics, on the one hand  (e.g. Meinel \& Brandenburg
1990, Moss {\em et al.} 1992, Ossendrijver \& Hoyng 1996, Ossendrijver, Hoyng
\& Schmitt 1996) and on the signatures and scalings of various
mechanisms of intermittency on the other.

\end{enumerate}

These issues raise a number of interesting questions. Is, for example,
the intermittency hypothesis operationally decidable at present?
Will it be operationally decidable in
foreseeable future?

In this connection it is worth bearing in mind that
some types of intermittency 
do possess signatures that are rather easily identifiable.
Nevertheless, we believe 
the answer to these difficult questions can only be realistically
contemplated once a more clear picture has emerged of all the possible types
of intermittency that can occur in more realistic solar-type dynamo models
(and ultimately real dynamos) and once their precise signatures and scalings,
in presence of noise, have
been identified.

\acknowledgments

We would like to thank the organisers of this meeting for their kind
hospitality and for bringing about the opportunity for many fruitful
exchanges. We would also like to thank Peter Ashwin,
Axel Brandenburg, John Brooke, David
Moss, Ilkka Tuominen and Andrew Tworkowski for the work we have done together
and Edgar Knobloch, Steve Tobias, Alastair Rucklidge, Michael Proctor and
Nigel Weiss for many stimulating discussions.

\vspace{0.2cm}
EC is supported by grant BD/5708/95 -- PRAXIS XXI, JNICT. EC thanks the
Astronomy Unit at QMW for support to attend the conference. RT benefited
from PPARC UK Grant No. L39094. This research also benefited from the
EC Human Capital and Mobility (Networks) grant ``Late type stars:
activity, magnetism, turbulence'' No. ERBCHRXCT940483.



\begin{references}
\reference Ashwin, P., Covas, E.\ \& Tavakol, R., 1999, {\em Nonlinearity}, {\bf 9}, 563.
\reference Baliunas, S.\ L.\ \& Soon, W., 1995, {\em Astrophy.\ J.}, {\bf 450}, 896.
\reference Beer, J.\ {\em et al.}, 1990, {\em Nature} {\bf 347}, 164.
\reference Beer, J.\ {\em et al.}, 1994a, in J.\ M.\ Pap, C.\ Fr\"ohlich, H.\ S.\ Hudson \& S.\ K.\ Solaski (eds.), {\it The Sun as a Variable Star: Solar and Stellar Irradiance Variations}, Cambridge University Press, Cambridge, p.\ 291.
\reference Beer, J.\ {\em et al.}, 1994b, in E.\ Nesme-Ribes (ed.), {\em The Solar Engine and its Influence on Terrestial Atmosphere and Climate}, Springer-Verlag, Berlin, p.\ 221.
\reference Brandenburg, A., 1999a, in: {\it Theory of Black Hole Accretion Discs}, eds.\ M.\ A.\ Abramowicz, G.\ Bj\"ornsson \& J.\ E.\ Pringle, Cambridge University Press.
\reference Brandenburg, A., 1999b, in: {\it Helicity and Dynamos}, eds.\ A.\ A.\ Pevtsov, American Geophysical Union, Florida.
\reference Brandenburg, A., Jennings R.\ L., Nordlund \AA., Rieutord M., Stein R.\ F., Tuominen, I., 1996, {\em JFM}, {\bf 306}, 325.
\reference Brandenburg, A., Nordlund, \AA., Stein, R.\ F., Torkelsson, U., 1995, {\em ApJ}, {\bf 446}, 741.
\reference Brooke, J.\ M., 1997, {\em Europhysics Letters} {\bf 37}, 3.
\reference Brooke, J.\ M., Pelt, J., Tavakol, R.\ \& Tworkowski, A., 1998, {\em A\&A} {\bf 332}, 339.
\reference Bussac, M.\ N.\ \& Meunier,C., 1982, {\em J.\ de Phys.}, {\bf 43}, 585.
\reference Butler, C.\ J.\ \& Johnston, D.\ J., 1996, {\em J.\ Atmospheric Terrest.\ Phys.}, {\bf 58}, 1657.
\reference Cattaneo, F., Hughes, D.\ W.\ \& Weiss, N.\ O., 1991, {\em MNRAS}, {\bf 253}, 479.
\reference Cattaneo, F.\ \& Hughes, D.\ W., 1996, {\em Phys.\ Rev.\ E} {\bf 54}, 4532.
\reference Covas, E., Tworkowski, A., Brandenburg, A.\ \& Tavakol, R., 1997a, {\em A\&A} {\bf 317}, 610.
\reference Covas, E., Tworkowski, A., Tavakol, R.\ \& Brandenburg, A., 1997b, {\em Solar Physics} {\bf 172}, 3.
\reference Covas, E., Ashwin, P.\ \& Tavakol, R., 1997c, {\em Physical Review E} {\bf 56}, 6451.
\reference Covas, E.\ \& Tavakol, R., 1997, {\em Physical Review E} {\bf 55}, 6641.
\reference Covas, E.\ \& Tavakol, R., 1998, Proceedings of the 5th International Workshop ``Planetary and Cosmic Dynamos'', Trest, Czech Republic, Studia Geophysica et Geodaetica, 42.
\reference Covas, E.\ \& Tavakol, R., 1999, {\it Multiple forms of intermittency in PDE dynamo models}, in preparation.
\reference Covas, E., Tavakol, R., Tworkowski, A., Brandenburg, A., Brooke, J.\ M.\ \& Moss, D., 1999a, {\em A\&A}, in press. Preprint available at web address \linebreak {\it http://www.maths.qmw.ac.uk/$\sim$eoc}.
\reference Covas, E., Tavakol, R., Ashwin, P., Tworkowski, A.\ \& Brooke, J.\ M., 1999b, submitted to {\em Phys.\ Lett.\ A}. Preprint available at web address \linebreak {\it http://www.maths.qmw.ac.uk/$\sim$eoc}.
\reference Eddy, J.\ A., 1976, {\em Science}, {\bf 192}, 1189.
\reference Feudel, W.\ Jansen, \& J.\ Kurths, 1993, {\em Int.\ J.\ of Bifurcation and Chaos} {\bf 3}, 131.
\reference Foukal, P.\ V., 1990, {\em Solar Astrophysics}, Wiley Interscience, New York.
\reference Friis-Christensen, E.\ \& Lassen, K., 1991, Science {\bf 254}, 698.
\reference Gilman, P.\ A., 1983, {\em ApJ.\ Suppl.}, {\bf 53}, 243.
\reference Gough, D., 1990, {\em Phil.\ Trans.\ R.\ Soc.\ Lond.} {\bf A330}, 627.
\reference Grebogi, C., Ott, E., Romeiras, F., \& Yorke, J.A., 1987, {\em Phys.\ Rev.\ A.}, {\bf 36}, 5365.
\reference Grebogi, C., Ott, E., \& Yorke, J.A., 1982, Phys. Rev Lett, {\bf 48}, 1507
\reference Hawley J.F., Gammie C.F., Balbus S.A., 1996, ApJ {\bf 464}, 690
\reference Hoyt, D.\ V.\ \& Schatten, K.\ H., 1996, {\em Solar Phys.} {\bf 165}, 181.
\reference Jones, C.\ A., Weiss N.O., Cattaneo F., 1985, {\em Physica 14D}, 161
\reference Knobloch, E., Tobias, S.\ M.\ \& Weiss, N.\ O., 1998, {\em MNRAS}, {\bf 297}, 1123.
\reference Krause, F.\ \& R\"adler, K.-H., 1980, {\em Mean Field Magnetohydrodynamics and Dynamo Theory}, Pergamon, Oxford.
\reference Lean, J., 1994, in E.\ Nesme-Ribes (ed.), {\it The Solar Engine and its Influence on Terrestial Atmosphere and Climate}, Springer-Verlag, Berlin, p.\ 163.
\reference Meinel, R.\ \& Brandenburg, A., 1990,{\em A\&A }{\bf 238}, 369.
\reference Moss, D., Barker D.M., Brandenburg A., Tuominen I., 1995,{\em A\&A }294, 155
\reference Moss, D., Brandenburg, A., Tavakol, R.\ \& Tuominen, I., 1992,{\em A\&A }{\ bf 265}, 843.
\reference Nordlund, \AA., Brandenburg, A., Jennings, R.\ L., Rieutord, M., Ruokolainen, J., Stein, R.\ F.\ \& Tuominen I., 1992, {\em ApJ}, {\bf 392}, 647
\reference O'Brien, S.\ R., Mayewsky, P.\ A., Meeker, L.\ D., Meese, D.\ A., Twickler, M.\ S.\ \& Whitlow, S.\ I., 1995, {\em Science}, {\bf 270}, 1962.
\reference Ossendrijver, A.\ J.\ H., Hoyng, P.\ \& Schmitt, D., 1996,{\em A\&A }{\bf 313}, 938.
\reference Ossendrijver, A.\ J.\ H.\ \& Hoyng, P., 1996,{\em A\&A }{\bf 313}, 959.
\reference Ott, E., {\em Chaos in Dynamic Systems}, 1993, Cambridge Press, Cambridge
\reference Platt, M., Spiegel, E.\ \& Tresser, C., 1993a, {\em Phys.\ Rev.\ Lett.}, {\bf 70}, 279.
\reference Pomeau, Y.\ \& Manneville, P., 1980, {\em Commun.\ Math.\ Phys.}, {\bf 74}, 189.
\reference Ribes, J.\ C.\ \& Nesme-Ribes, E., 1993, {\em A\&A} {\bf 276}, 549.
\reference Richter, R., Kittel, A., Heinz, G., Fl\"atgen, G., Peinke, J.\ \& Parisi, J., 1994, {\em Phys.\ Rev.\ B} {\bf 49}, 8738.
\reference Ruzmaikin, A.\ A., 1981, {\em Comm.\ Astrophys.}, {\bf 9}, 88.
\reference Schmalz, S.\ \& Stix, M., 1991, {\em A\&A} {\bf 245}, 654.
\reference Schmitt, D., Sch\"ussler, M., \& Ferriz-Mas, A., 1996, A\&A, {\bf 311}, L1.
\reference Spiegel, E., Platt, N.\ \& Tresser, C., 1993b, {\em Geophys.\ and Astrophys.\ Fluid Dyn.}, {\bf 73}, 146.
\reference Spiegel, E.A.\ 1994, in Proctor M.R.E., Gilbert A.D., eds, {\it Lectures on Solar and Planetary Dynamos}, Cambridge Univ.\ Press, Cambridge.
\reference Spiegel, in {\it Chaos in Astrophysics}, edited by J.\ R.\ Butcher, J.\ Perdang, \& E.\ A.\ Spiegel (Reidel, Dordrecht, 1985).
\reference Stuiver, M., Grootes, P.\ M.\ \& Braziunas, T.\ F., 1995, {\em Quarternary Res.} {\bf 44}, 341.
\reference Stuiver, M.\ \& Braziunas, T.\ F., 1988, in F.\ R.\ Stephenson \& A.\ W.\ Wolfendale (eds.), {\it Secular Solar and Geomagnetic Variations in the Last 10 000 Years}, Kluwer Academic Publishers, Dordrecht, Holland, p.\ 245.
\reference Stuiver, M.\ \& Braziunas, T.\ F., 1989, {\em Nature} {\bf 338}, 405.
\reference Stuiver, M.\ \& Quay, P.\ D., 1980, {\em Science}, {\bf 207}, 19.
\reference Tavakol, R., 1978, {\em Nature}, {\bf 276}, 802.
\reference Tobias, S.\ M., 1998, {\em MNRAS}, {\bf 296}, 653.
\reference Tobias, S.\ M., Weiss, N.O. \& Kirk, V., 1995, {\em MNRAS}, {\bf 273}, 1150.
\reference Tworkowski, A., Tavakol, R., Brandenburg, A., Brooke, J.\ M., Moss, D.\ \& Tuominen I., 1998, {\em MNRAS}, {\bf 296}, 287.
\reference Weiss, N.\ O., Cattaneo, F., Jones, C. A., 1984, {\em Geophys.\ Astrophys.\ Fluid Dyn.}, {\bf 30}, 305.
\reference Weiss, N.\ O., in {\em Lectures on Solar and Planetary Dynamos}, edited by Proctor, M.R.E. and Gilbert, A.D., Cambridge University Press, Cambridge (1994)
\reference Weiss, N.\ O.\ \& Tobias, S.\ M., in Solar and Heliospheric Plasma Physics, ed. G.\ M.\ Simnett, C.\ E.\ Alissandrakis \& L.\ Vlahos, 25, Springer, Berlin, 1997.
\reference Weiss, N.\ O.\, 1990, {\em Phil.\ Trans.\ R.\ Soc.\ Lond.}, {\bf A330}, 617.
\reference White, W.\ B., Lean, J., Cayan, D.\ \& Dettinger, M.\ D., 1997, {\em J.\ Geophys.\ Res.}, {\bf 102}, 3255.
\reference Wilson, P.\ R., 1994, {\it Solar and Stellar Activity Cycles}, Cambridge University Press, Cambridge.
\reference Zeldovich, Ya.\ B., Ruzmaikin, A.\ A.\ \& Sokoloff, D.\ D., 1983, {\it Magnetic Fields in Astrophysics}, Gordon and Breach, New York.
\end{references}
\end{document}